\documentclass[twocolumn,showpacs,secnumarabic,amssymb,nobibnotes,aps,prl]{revtex4}

\usepackage{amsmath}
\usepackage{latexsym}
\usepackage{float}
\usepackage{amssymb}
\usepackage{graphicx}
\usepackage{textcomp}
\usepackage{hyperref}
\begin{document}

\title{Does Young's equation hold on the nanoscale? A Monte Carlo test for the binary Lennard-Jones fluid}
\author{Subir K. Das$^{1,2}$ and Kurt Binder$^{2}$}
\affiliation{$^1$ Theoretical Sciences Unit, Jawaharlal Nehru
Centre for Advanced Scientific Research, Jakkur, Bangalore,
560064, India\\
$^2$ Institut f\"ur Physik, Johannes Gutenberg-Universit\"at, Staudinger Weg 7, D55099 Mainz, Germany\\
 }

\date{\today}

\begin{abstract}
When a phase-separated binary ($A+B$) mixture is exposed to a wall,
that preferentially attracts one of the components, interfaces
between A-rich and B-rich domains in general meet the wall making a
contact angle $\theta$. Young's equation describes this angle in
terms of a balance between the $A-B$ interfacial tension
$\gamma_{AB}$ and the surface tensions $\gamma_{wA}$,
$\gamma_{wB}$ between, respectively, the $A$- and $B$-rich phases and the wall,
$\gamma _{AB} \cos \theta =\gamma_{wA}-\gamma_{wB}$. By Monte
Carlo simulations of bridges, formed by one of the components in a
binary Lennard-Jones liquid, connecting the two
walls of a nanoscopic slit pore,
$\theta$ is estimated from the inclination of the
interfaces, as a function of the wall-fluid interaction strength. The information on the surface
tensions $\gamma_{wA}$, $\gamma_{wB}$ are obtained independently
from a new thermodynamic integration method, while $\gamma_{AB}$
is found from the finite-size scaling analysis of the
concentration distribution function. We show that Young's equation
describes the contact angles of the actual nanoscale interfaces
for this model rather accurately and location of
the (first order) wetting transition is estimated.
\end{abstract}
\maketitle

{\bf Introduction and Motivation}:
Pure fluids and fluid mixtures confined to nanopores are of substantial
recent research interest due to various applications, e.g., extraction of oil or
gas from porous natural rocks, use as ``molecular
sieves'' to separate fluids, and various applications in
nanofluidics such as ``lab on a chip''-devices and preparation of
materials in nanotechnology, etc. \cite{1,2,3,4,5,6}. At the same
time, the interplay between finite-size effects, surface effects due to
the pore walls, and interfacial phenomena pose challenging
problems to the theoretical understanding on the basis of
statistical mechanics \cite{7,8,9,10}. A fundamental concept in
this context is the contact angle $\theta$ 
with which an interface between coexisting (bulk) phases
in a phase-separated system
meets a wall under conditions of incomplete wetting
\cite{7,8,9,10,11,12,13}. In the limit of macroscopically large
droplets, for which the droplet radius is many orders of magnitude
larger than any interfacial width, one believes that 
$\theta$ can be expressed in terms of a balance between
appropriate interfacial tensions via Young's equation \cite{14}. For the case of a
binary ($A+B$) fluid mixture undergoing phase separation in the bulk
into $A$-rich and $B$-rich coexisting phases, this equation 
reads
\begin{equation} \label{eq1}
\gamma_{AB} \cos \theta=\gamma _{w A} - \gamma _{wB},
\end{equation}
where $\gamma_{AB}$ is the interfacial excess free energy due to
an infinitely extended flat interface between the coexisting
bulk phases. In Eq.~(\ref{eq1}), the excess free energies of the A-rich and B-rich
phases due to the wall are denoted by $\gamma_{wA}$ and $\gamma_{wB}$,
respectively.

While numerous techniques exist for precise experimental
measurements of both the contact angle of macroscopic droplets and
the interfacial tension $\gamma_{AB}$, it is challenging
to measure $\gamma_{wA}$, $\gamma_{wB}$ independently with
sufficient precision. Thus a test of Eq.~(\ref{eq1}) even for
macroscopic droplets is a very difficult task and furthermore,
often hampered by substrate inhomogeneities causing contact angle
hysteresis \cite{1,11}. For pores in the nanoscale size range,
the problem is much harder, since the atomistically diffuse
character of the interfaces can no longer be neglected, and
of course, in many circumstances interfaces are curved where the
understanding of how the interfacial tension $\gamma_{AB}(R)$ varies
with the radius of curvature is still incomplete \cite{15,16,17,18}.
An additional complication already comes into play for sessile
mesoscale droplets attached to walls: then Eq.~(\ref{eq1}) is
modified by a line tension ($\tau$) contribution from the
circumference length $2 \pi r$ of the sphere-cap shaped droplet right at the
wall \cite{19,20},
\begin{equation} \label{eq2}
\gamma _{AB} \cos \theta=\gamma _{wA} - \gamma_{wB} + \tau/r.
\end{equation}

Estimation of the magnitude of the line tension $\tau$ is another
problem very controversially discussed in the literature
\cite{21,22,23,24,25}. Indeed the proper identification of the line
tension effects is subtle \cite{26}. Despite a large experimental
and theoretical activity 
\cite{27,28,29,30,31,32,33,34,35,36,37} for studying capillary
condensation, liquid bridges in slit pores, etc., typically the
information obtained does not yield both the contact angle and all
the excess free energies in Eqs.~(\ref{eq1}) and (\ref{eq2}) with the
notable exception of recent simulation studies of the
Ising lattice-gas model \cite{36,37}.

The purpose of the present work hence is to provide a first
complete test of the validity of Eqs.~(\ref{eq1}) and (\ref{eq2}) in
the nano-scale via Monte Carlo simulations of an off-lattice model
of a fluid binary mixture. We shall study the structure of liquid
bridges of A-rich phases in a B-rich background for various
choices of the interactions between the particles and the walls of
a nanoscopic slit pore and extract effective contact angles
$\theta_{\rm eff}$ to describe the inclined interfaces that occur.
In addition, we compute $\gamma_{AB}$ for bulk interfaces and
develop a new method to compute $\gamma_{wA}$, $\gamma_{wB}$ for pure
semi-infinite phases exposed to corresponding walls. Extending the
approach of Winter {\it et al.} \cite{36,37}, we are also able to derive the line
tension $\tau$ as function of the contact angle $\theta$.

{\bf Model and Simulation Technique}:
Let us consider a binary ($A+B$) fluid consisting of $N$ point particles, labelled by
index $i$ at positions $\vec{r}_i$, in a box of linear dimensions
$L \times L \times D$, with periodic boundary conditions in $x$-
and $y$- directions and having impenetrable walls of area $L \times L$ each at
$z=0$ and $z=D$. The particles interact via pairwise potentials $u(r_{ij}
=|\vec{r_i}-\vec{r}_j|)$, constructed from the full
Lennard-Jones (LJ) potential $\phi_{LJ} (r_{ij}) = 4
\varepsilon_{\alpha \beta} [(\sigma_{\alpha \beta}/r_{ij})^{12} -
(\sigma_{\alpha \beta}/ r_{ij})^6]$, $\alpha, \beta \in A,B$,
as \cite{38,39}
\begin{equation} \label{eq3}
u(r_{ij}\leq r_c)= \phi_{LJ} (r_{ij}) - \phi_{LJ} (r_c) - (r_{ij} -r_c)
\frac{d \phi_{LJ}}{d r_{ij}} \mid_{r_{ij}=r_c},
\end{equation}
while $u(r_{ij} \geq r_c)=0$. This ensures that both the potential and
the force are continuous at the cut-off distance $r_{ij}=r_c$. 
The potential parameters
are chosen such that the mixture is symmetric:
$\sigma_{AA}=\sigma_{BB}=\sigma_{AB}=\sigma$,
$\varepsilon_{AA}=\varepsilon_{BB}=2\varepsilon_{AB}=\varepsilon$
and $r_c$ was set at $2.5\sigma$.
Working at a reduced
density $\rho^*=1$, where $\rho^*=\rho \sigma^3=N \sigma^3/V$,
neither crystallization nor the vapor-liquid transition is a
problem, while unmixing occurs for \cite{38,39} $T<T_c=1.4230 \pm
0.005$ (unit of $T$ is chosen such that $\varepsilon/k_B \equiv 1$).
Working at $T=1.0$, the bulk A-rich (B-rich) phases are almost
pure, $x_{A}^{\rm
coex}=(N_A/N)_{\rm coex}\approx 0.97$ ($x_B^{\rm coex}=0.03$) \cite{18}.

We choose ``antisymmetric'' interactions of the walls with the fluid particles such that
also in the thin film geometry phase coexistence between A-rich
and B-rich phases occurs at chemical potential difference
$\Delta\mu=0$ between A- and B-particles, as in the bulk
\cite{38,39}. Specifically, for the A particles we take a wall
potential
\begin{eqnarray} \label{eq4}
u_A (z) = \frac{2 \pi \rho^*}{3}\times\quad \quad\quad\quad\quad\quad\quad\quad\quad\quad\quad\quad
\quad\quad\quad \nonumber\\
\Big\{ \varepsilon_r
\Big[\Big(\frac{\sigma}{ z + \delta}\Big)^9 + \Big(\frac{\sigma}{D
+ \delta -z}\Big)^9 \Big] - \varepsilon_a \Big(\frac{\sigma}{z +
\delta}\Big)^3 \Big\},
\end{eqnarray}
and for the B-particles
\begin{eqnarray} \label{eq5}
u_B(z) =\frac{2 \pi \rho^*}{3}\times \quad \quad\quad\quad\quad\quad\quad\quad\quad\quad\quad\quad 
\quad\quad\quad\nonumber\\
\Big\{ \varepsilon_r
\Big[\Big(\frac{\sigma}{z+ \delta}\Big)^9 + \Big(\frac{\sigma}{D+
\delta-z} \Big)^9 \Big] - \varepsilon_a \Big(\frac{\sigma}{D+
\delta-z}\Big)^3 \Big\},
\end{eqnarray}
where $0 \leq z \leq D$ is the coordinate perpendicular to the
walls, and the offset $\delta = \sigma/2$. Both the walls exert the
same repulsive potential (of strength
$\varepsilon_r=\varepsilon/2)$ on both kinds of particles. This
potential can be thought of as resulting from integration over LJ
repulsions from atoms residing in a semi-infinite wall. An
attractive interaction (of variable strength $\varepsilon_a$) acts
only on the A-particles from the wall at $z=0$, while the same
acts only on the B particles from the wall at $z=D$.

For direct measurement of $\theta$ from the inclination of $A-B$
interface,
Monte Carlo (MC) simulations are carried out in the
canonical ensemble where in addition to the standard
particle displacement moves, exchange between randomly
chosen particle pairs was also tried.
For the displacements, a trial
shift of a cartesian coordinate of a randomly selected particle in
the range $[-\sigma/20$, $+ \sigma/20]$ around its old position is
chosen and the standard Metropolis criterion applied. 
On the other hand, thermodynamic integration (see below) was performed
by using the raw data obtained from MC simulation in 
the semi-grand-canonical (SGMC) ensemble. In the latter,
after every 10
displacement steps per particle $N/10$ particles are randomly
chosen in succession and an attempted identity switch 
($A \rightarrow B \rightarrow A$) is made (for
more technical details, see \cite{38,39}). In addition,
successive umbrella sampling was performed using SGMC to 
obtain interfacial free energies of the co-existing phases.

\begin{figure}
\centering
\includegraphics*[width=0.4\textwidth]{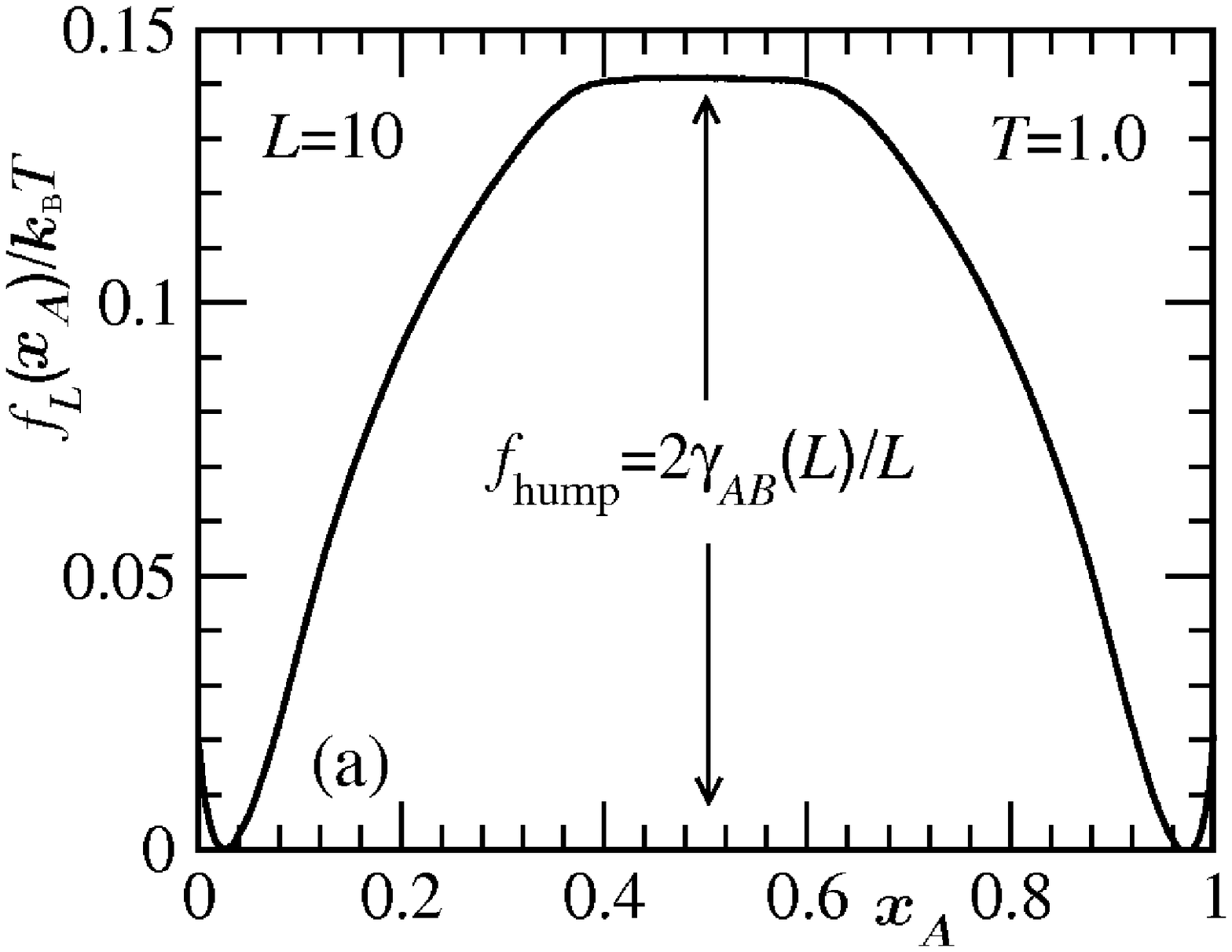}
\includegraphics*[width=0.4\textwidth]{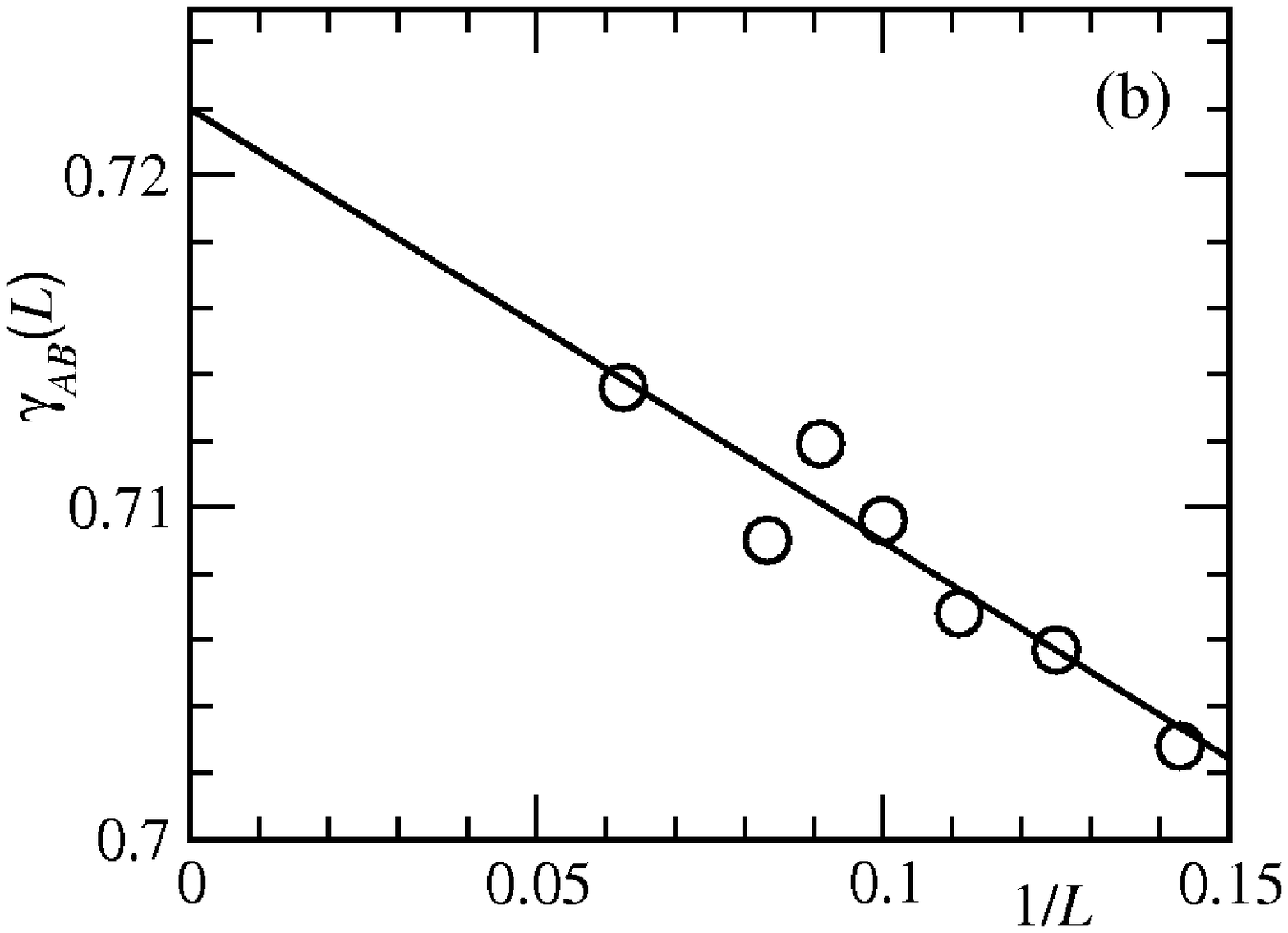}
\caption{(a) Effective free energy density $f_L(x_A,T)$ of the
symmetric LJ mixture at $T=1.0$ for a box of linear dimension
$L=10$ (in units of $\sigma$) plotted vs. $x_A$. The
estimation of the size-dependent interfacial tension
$\gamma_{AB}(L)$ is indicated. (b) Extrapolation of
$\gamma_{AB}(L)$ as function of $1/L$.} \label{fig1}
\end{figure}

As a first step, we estimate $\gamma_{AB}$ from the finite-size
extrapolation of the distribution $P_L(x_A)$ 
($x_A=N_A/N$, the mole fraction of A-particles) using a $L \times
L \times L $ geometry, with periodic boundary conditions
in all $x$-, $y$- and $z$- directions. Applying
successive umbrella sampling \cite{40}, the effective excess free energy
density $f_L(x_A,T)$ is found from $f_L(x_A, T)= -(1/V) \ln
[P_{\Delta \mu NVT} (x_A)/P_{\Delta \mu NVT} (x_A^{\rm coex})]$.
As is well-known \cite{18,37,41}, the flat maximum, $f_{\rm hump}$, in $f_L(x_A,T)$,
near $x_A=0.5$ in Fig.~\ref{fig1},
is due to the formation of two
planar $L \times L$ interfaces in the simulation box,
separating a slab of A-rich phase from the B-rich phase (or vice
versa). Thus this barrier is given by $2
\gamma_{AB} (L)/L$ and the extrapolation of such data towards $L
\rightarrow \infty$ yields $\gamma_{AB}=0.722
\pm 0.002$, as demonstrated in Fig.~\ref{fig1}(b).

\begin{figure}
\centering
\includegraphics*[width=0.4\textwidth]{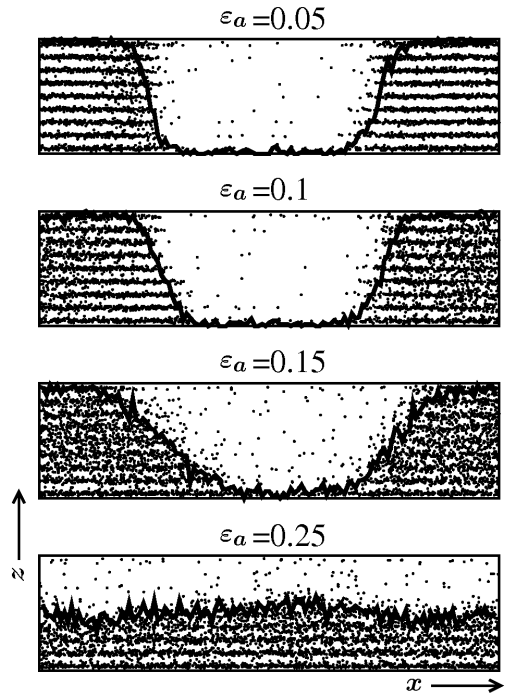}
\caption{Snapshot pictures of the configurations of the A-
particles (dots) in a $L \times L \times D$ system with $L=32$,
$D=8$, projected onto the $xz$-plane, for the case of interfaces
running along the $y$-direction. B-particles are not shown.
Four choices of attractive wall-particle interaction strength
$\varepsilon_a$ are included, as indicated. The
instantaneous interface position (averaged over the $y$
coordinate) is highlighted by thick lines.} \label{fig2}
\end{figure}

\begin{figure}
\centering
\includegraphics*[width=0.4\textwidth]{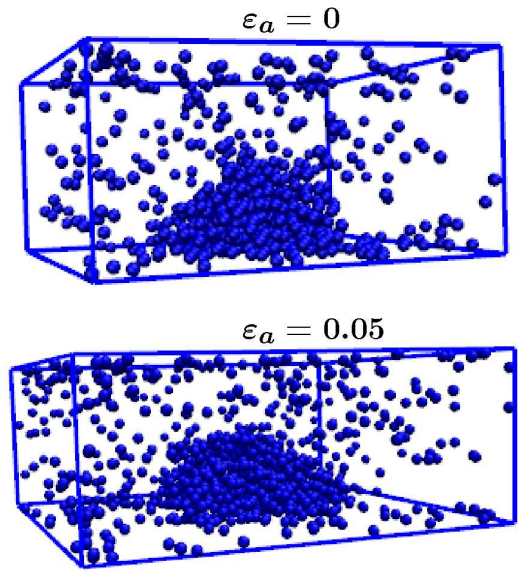}
\caption{
Three-dimensional
snapshots of $A-particle$ droplet configurations for the cases $x_A =0.12$,
$\varepsilon_a=0$ ($D=12$, $L=24$; upper part) and $x_a=0.12$,
$\varepsilon_a=0.05$ ($D=12$, $L=30$; lower part).} \label{fig3}
\end{figure}

{\bf Interfaces in nano-confinement}:
Using the model in a $L \times L \times D$ geometry, applying
the wall potentials (\ref{eq4}) and (\ref{eq5}) and starting with 
initial configurations prepared such that two $A-B$ interfaces meet the
walls vertically, we
stabilize two-phase configurations of the system
via MC simulation in canonical ensemble
for varying strengths of attractive wall potential $\varepsilon_a$.
For $x_A=0.5$ and $\varepsilon_a \leq0.2$ we find slab-like
configurations of the B-rich phase as seen in Fig.~\ref{fig2},
where the interfaces are represented by thick lines. These
interfaces are perpendicular to the $xy$-plane only for
$\varepsilon_a=0$ when the contact angle $\theta= \pi/2$ by
symmetry of our model. For $\varepsilon_a > 0$ they are
inclined relative to the $xy$-plane. We extract an effective
contact angle $\theta _{\rm eff}$ from the average slope of these
interfaces, measured from the central part of the film, excluding
regions of width $2 \sigma$ at both walls. For
$\varepsilon_a=0.25$, wetting has already occurred ($\theta_{\rm eff}
= 0$) and so, instead of a slab with two interfaces, we now
find a single interface, separating the $A$-rich domain 
near the bottom wall that prefers $A$ from $B$-rich domains near the
top wall preferring $B$. One consequence of nano-confinement that
one can see in all cases is a pronounced layering of the total
density parallel to the confining walls and extending throughout
the thin film.
If one chooses suitable off-critical concentrations, e.g.,
$x_A=0.12$, one can observe wall-attached droplets 
shown in Fig.~\ref{fig3}, similar to the
Ising case \cite{36,37}. The analysis
of such states when obtained via umbrella sampling method,
thus containing the information of $f_L$,
allows the estimation of the line tension, when the
appropriate contact angle is independently known.

\begin{figure}
\centering
\includegraphics*[width=0.4\textwidth]{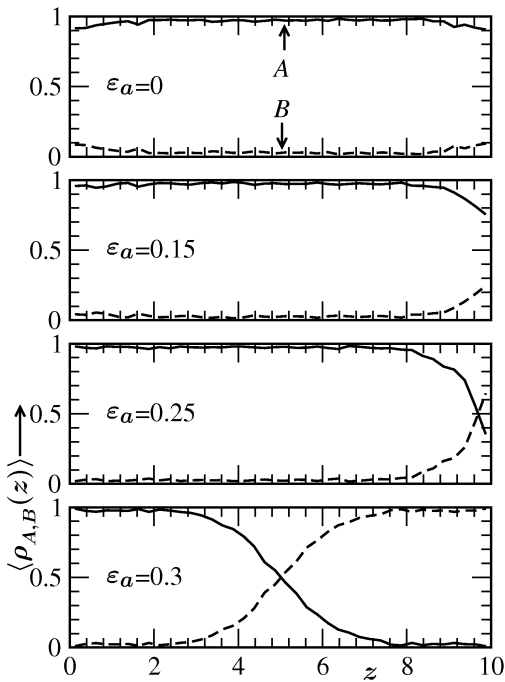}
\caption{Density profiles $\langle \rho_A(z)\rangle$,
$\langle \rho_B(z)\rangle$, obtained from the semi-grand-canonical
MC simulation in a $L\times L\times D$ geometry averaging over $x$ and $y$
coordinates,
plotted vs. $z$
for a film of thickness $D=10$. Three top panels
($\varepsilon_a=0$, $0.15$ and $0.25$)
correspond to the A-rich phases.
The panel at the
bottom with $\varepsilon_a=0.3$ shows the profiles for a state
where unbinding of the interface from the wall at $z=D$ has occurred,
with the delocalized interface now being at $z=D/2$.} \label{fig4}
\end{figure}

{\bf Thermodynamic
Integration and Bulk Contact Angle}:
We consider a film of large but finite thickness in the
semi-grand-canonical ensemble with $\Delta \mu=0$ and vary
$\varepsilon_a$ being in the regime of incomplete wetting, where the
system is either in the pure A-rich phase
or in the pure B-rich phase, prepared via SGMC simulation. From
the density profiles of such systems, as shown in Fig.~\ref{fig4},
where the three upper panels correspond to $A$-rich phases
(note that in this case only a slight enhancement of $B$ particles could
be seen at the wall preferring the latter),
one can extract information on
$\gamma_{wA}-\gamma_{wB}$ by a thermodynamic integration method.
To derive this, we first express the total free energy $F$ of
the film via the partition function as a configurational
integral 
\begin{eqnarray} \label{eq6}
&& F=-k_BT \ln Z\nonumber\\
&&~~ =-k_BT \ln \int\limits d \vec{X} \exp \Big\{-\beta
\mathcal{H}_b (\vec{X})-\beta \mathcal{H}^r_w (\vec{X}) \nonumber\\
&& + \beta \varepsilon_a L^2 ( 2 \pi \rho^*/3)\times\nonumber\\
&&\Big[ \int\limits_0^D
\langle \rho_A(z)\rangle \Big(\frac{\sigma}{z + \delta}\Big)^3 dz +\nonumber\\
&&\int\limits_0^D \langle \rho_B (z)\rangle \Big(\frac{\sigma}{D+\delta-z}\Big)^3
dz \Big]\Big\}.
\end{eqnarray}
Here  $\beta =1/k_BT$ and  $\vec{X}$ includes all configurational degrees of freedom of
the system. For the sake of brevity, neither the bulk energy
$\mathcal{H}_b(\vec{X})$ nor the energy $\mathcal{H}^r_w(\vec{X})$
due to the repulsive part of the wall
potential has been explicitly written
down. The density profiles $\langle \rho_{A,B}(z)\rangle$
are averaged over $x$ and $y$ coordinates.
We now take a derivative with respect to $\varepsilon_a$,
which singles out the derivatives of the wall free energies $f_s$
at $z=0$ and $z=D$, respectively ($f_s$ is normalized per unit
area)
\begin{equation} \label{eq7}
\Big(\frac{\partial f_s^{(z=0)}}{\partial \varepsilon_a}\Big)_T
=\Big(2 \pi \rho^* / 3 \Big) \int\limits_0^D dz \Big(
\frac{\sigma}{z +\delta} \Big)^3 \langle \rho_A (z )\rangle,
\end{equation}
\begin{eqnarray} \label{eq8}
\Big(\frac{\partial f_s^{(z=D)}}{\partial \varepsilon_a } \Big)_T &=&
(2 \pi \rho^* /3 )\times \nonumber\\
&& \int\limits_0^D dz \Big(\frac{\sigma}{D+
\delta -z} \Big)^3 \langle \rho _B (z)\rangle.
\end{eqnarray}
These equations are readily integrated to yield
\begin{eqnarray} \label{eq9}
&&f_s^{(z=0)} (\varepsilon_a)=f_s^{(z=0)} (0) + ( 2 \pi \rho^*/3) \times \nonumber\\
&& \int\limits_0^{\varepsilon_a} d \varepsilon'_a \int\limits_0^D dz
\Big(\frac{\sigma}{z + \delta} \Big)^3
<\rho_A (\varepsilon_a', z)>,
\end{eqnarray}
\begin{eqnarray} \label{eq10}
&&f_s^{(z=D)} (\varepsilon_a) = f_s^{(z=D)} (0) + (2 \pi \rho^*/3) \times \nonumber\\
&& \int\limits_0^{\varepsilon_a} d \varepsilon'_a \int\limits_0^D dz
 \Big(\frac{\sigma}{D + \delta - z} \Big)^3
<\rho_B (\varepsilon'_a, z)>.
\end{eqnarray}

Noting that surface free energies of bulk B-rich phases are
related to those of the A-rich phases simply by symmetry:
\begin{equation} \label{eq11}
f_s^{(z=0)} (\varepsilon_a) \mid_{{B-{\rm rich \,  phase}}}
=f_s^{(z=D)} (\varepsilon_a) \mid_{{A-{\rm rich \, phase}}},
\end{equation}
we immediately obtain the desired difference
\begin{eqnarray} \label{eq12}
&& \gamma_{wA} - \gamma_{wB}  \equiv  f_s^{(z=0)}
(\varepsilon_a)\mid_{A-{\rm rich \, phase}} - \nonumber\\
&& f_s^{(z=0)}
(\varepsilon_a) \mid_{B-{\rm rich \, phase}} \nonumber\\
&& = (2 \pi \rho^* /3) \int\limits_0^{\varepsilon_a} d
\varepsilon'_a \int\limits_0^D dz \times \nonumber\\
&& \Big[ \langle \rho_A
(\varepsilon'_a, z) \rangle _{_{A-{\rm rich}}} \Big(\frac{\sigma}{z + \delta} \Big)^3 - \nonumber\\
&&  \langle \rho_B (\varepsilon'_a, z) \rangle _{_{A-{\rm rich}}} \Big(\frac{\sigma}
{D+ \delta -z}\Big)^3 \Big].\nonumber\\
&&
\end{eqnarray}
In Eq.~(\ref{eq12}) both the profiles $\langle
\rho_A(\varepsilon'_a,z) \rangle$ and $\langle
\rho_B(\varepsilon'_a, z) \rangle$ are sampled, without loss of
generality, in the A-rich phase, as shown in Fig.~\ref{fig4}.

\begin{figure}
\centering
\includegraphics*[width=0.4\textwidth]{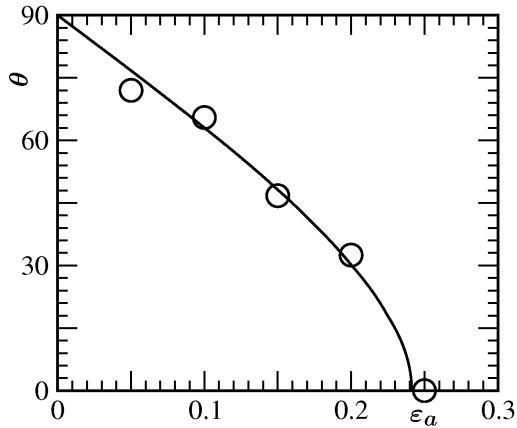}
\caption{The full curve is a plot of contact angle $\theta$ vs. $\varepsilon_a$, as
obtained from Young's equation with the information about $\gamma_{wA}-\gamma_{wB}$
being provided
from thermodynamic integration, Eq.~(\ref{eq12}).
Circles show corresponding observations from
Fig.~\ref{fig2}.} \label{fig5}
\end{figure}

Given the knowledge of both the difference $\gamma_{wA} -
\gamma_{wB}$ (as function of $\varepsilon_a$) and of
$\gamma_{AB}$, we can predict the dependence of the contact angle
$\theta$ on $\varepsilon_a $, as shown in Fig.~\ref{fig5}. Following this we obtain
the wetting transition ($\theta=0$) at
$\varepsilon_a=0.240\pm 0.005$. Note also that in our model, for
$\varepsilon_a=0$ we must have $\theta = \pi/2$ due to the full
symmetry between $A$ and $B$ that is present then. Interestingly, the
contact angles predicted from Young's equation, which refer to
quasi-macroscopic droplets, agree almost perfectly with the
contact angles, represented by circles in Fig.~\ref{fig5}, 
observed in a nanoscopically thin film
that is only $8$ Lennard-Jones diameters thick
(Fig.~\ref{fig2}). Thus Young's equation still is useful at the
nanoscale!

\begin{figure}
\centering
\includegraphics*[width=0.4\textwidth]{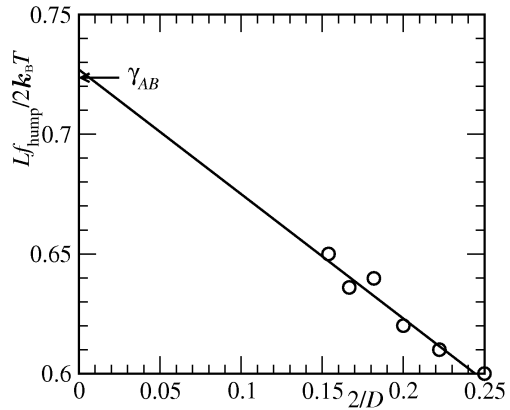}
\caption{Effective interface tension [estimated via
$Lf_{\rm hump}/(2k_BT)$] plotted vs. $2/D$, for systems of linear
dimensions $L \times L \times D$, with $L=30$ and using several
choices of $D$, for the case $\varepsilon_a=0$. Arrow shows the
estimate of $\gamma_{AB}$ from Fig.~\ref{fig1}(b). The straight line is a fit
to the form $\gamma_{AB} + 2\tau/D$, where both $\gamma_{AB}$ and $\tau$
have been used as adjustable parameters. Fitting gives $\tau=-0.52$ and
$\gamma_{AB}=0.725$ [note the agreement with the estimate from Fig.~\ref{fig1}(b)].}
\label{fig6}
\end{figure}

{\bf Estimation of the line tension}:
It would be wrong to conclude from this success of
Young's equation that the line tension does not play any role in
our model. In fact for the case $\varepsilon_a=0$ it is
relatively easy to estimate it
from the flat free energy maximum, $f_{\rm hump}$ (cf.
Fig.~\ref{fig1}(a)), for systems in $L \times L \times D$ thin film
geometry by varying $D$. Then one expects that $f_{\rm hump}$ has a
value $2 \gamma_{AB}/L + 4 \tau/(LD)$; hence a plot of $Lf_{\rm hump} / (
2k_BT)$ vs. $2/D$, as presented in Fig.~\ref{fig6}, 
should yield a straight line, where the
$y$-intercept is the already known interfacial tension
$\gamma_{AB}$, while the slope yields an estimate of $\tau$. In
this way we find $\tau \approx-0.52$ for $\varepsilon_a=0$.
Recall that by symmetry, $\theta = \pi/2$ for
$\varepsilon_a=0$, irrespective of the presence of the line
tension. Unfortunately, when $\varepsilon_a >0$ and the interfaces
are inclined (Fig.~\ref{fig2}), large statistical fluctuations
prevent us from using this method.
However, following Winter et al. \cite{36,37} we have analyzed the
excess free energy of sphere-cap shaped droplets. As
will be described elsewhere in detail \cite{42}, the absolute
magnitude of $\tau$, obtained from this analysis, decreases rapidly with decreasing contact
angle.

{\bf Summary}:
In this letter, we have described a study of the
contact angle of nanoscale fluid bridges, for a model of a
symmetric binary Lennard-Jones mixture, for the full range from
neutral walls up to the wetting transition. Supplying the
interfacial tensions entering Young's equation,
we observe that the quantitative agreement of the
contact angles thus obtained with the direct observation from 
inclined nanoscopic interfaces
in a $50:50$ composition is remarkable. While
$\gamma_{AB}$ and the line tension $\tau$ are extracted from
suitable system size dependences,  $\gamma_{wA} - \gamma_{wB}$ is
estimated via a new thermodynamic integration method. Clearly, the
symmetric Lennard-Jones mixture is a simple model system, but we
expect that our methods can be generalized to more complex models
of real fluids. In this way, practically relevant information to
guide the development of nanofluidic devices will come in reach.\\

\underline{Acknowledgement}: One of us (S.K.D.) is grateful to the
Deutsche Forschungsgemeinschaft (DFG) for partial support under
grant N$^o$ TR6/A5, and thanks the Institut f\"ur Physik, Mainz,
for hospitality during his research visits. We are grateful to J.
Horbach, M. Oettel and P. Virnau for stimulating discussions.


\end{document}